\documentclass[%
 reprint,
 amsmath,amssymb,
 aps,
pra,
floatfix,
]{revtex4-2}

\usepackage{graphicx}
\usepackage{dcolumn}
\usepackage{bm}
\usepackage{svg}
\usepackage[capitalise]{cleveref}

\newcommand{\ket}[1]{|#1 \rangle}

\newcommand{\ud}{\mathrm{d}}
\renewcommand{\b}{\hat{b}}
\newcommand{\bd}{\hat{b}^\dagger}
\newcommand{\s}{\mathrm{s}}
\newcommand{\p}{\mathrm{p}}
\newcommand{\qi}{\mathrm{q}}
\renewcommand{\a}{\hat{a}}
\newcommand{\ad}{\a^\dagger}
\newcommand{\A}{\hat{A}}
\newcommand{\Ad}{\A^\dagger}
\renewcommand{\i}{\mathrm{i}}
\newcommand{\m}{\hat{m}}

\newcommand{\md}{\hat{m}^\dagger}
\newcommand{\mean}[1]{\left\langle #1 \right\rangle}

\newcommand{\e}{\mathrm{e}}
\renewcommand{\vec}[1]{\mathbf{#1}}

\renewcommand{\th}{\mathrm{th}}
\newcommand{\q}{\hat{q}}

\newcommand{\hc}{\mathrm{h.c.}}

\renewcommand{\AA}{\hat{\mathcal{A}}}
\newcommand{\MM}{\hat{\mathcal{M}}}
\newcommand{\fR}{f_\mathrm{R}}
\newcommand{\tens}[1]{\underline{\mathbf{#1}}}

\begin{document}

\preprint{APS/123-QED}

\title{Raman effects in Quantum Frequency Conversion using Bragg Scattering}

\author{Mathias L. H. Korsgaard}
\author{Jacob G. Koefoed}%
\author{Karsten Rottwitt}
 \email{karo@dtu.dk}
\affiliation{%
 Department of Electrical and Photonics Engineering, Technical University of Denmark, 2800 Kongens Lyngby, Denmark
}%


\date{\today}

\begin{abstract}
We present a quantum-mechanical model that describes fiber-based frequency conversion by four-wave-mixing Bragg scattering in the presence of Raman interactions. In the case of continuous-wave pumps we find closed-form expressions for the conversion efficiency and photon statistics, characterized by the second-order correlation function. For pulsed pumps, we derive a highly general model based on Green functions, and provide a numerical solution method using a split-step scheme. In both cases, we find that noise from spontaneous Raman scattering can pose a serious challenge to this type of frequency conversion if the pumps are less than 30 THz from the quantum fields. However, this impact can be mitigated with crosspolarized pumps and on the anti-Stokes side, through cooling of the fiber.
\end{abstract}

\maketitle

\section{Introduction}
Quantum frequency conversion is a crucial component in future photon-based quantum communication and computing applications \cite{knillSchemeEfficientQuantum2001,divincenzoPhysicalImplementationQuantum2000a, laddQuantumComputers2010a, lukensFrequencyencodedPhotonicQubits2017}.
The ability to flexibly convert photons between wavelengths, without destroying the fragile quantum states \cite{maSinglePhotonFrequency2012,vanleentEntanglingSingleAtoms2022}, provides an interface between different parts of a quantum-optical system such as a photon source, a transmission fiber, or a quantum memory~\cite{maringPhotonicQuantumState2017}.

Four-wave mixing Bragg-scattering (FWM-BS) provides a platform free of additional noise~\cite{mcguinnessTheoryQuantumFrequency2011}, ideal for quantum frequency conversion.
Compared to the well-known processes of sum-frequency and difference-frequency generation \cite{wangQuantumFrequencyConversion2023a, hansenEfficientRobustSecondharmonic2023}, the two pump fields in FWM-BS provides increased flexibility~\cite{inoueTunableSelectiveWavelength1994}. 
In addition, FWM-BS is possible in optical fiber with efficient integration into an existing network~\cite{farsiLowNoiseQuantumFrequency2015a, clarkHighefficiencyFrequencyConversion2013a}.
This has proved to be a promising platform for frequency multiplexing of single-photons~\cite{joshiFrequencyMultiplexingQuasideterministic2018a}, and has furthermore been used to demonstrate quantum interference between spectral channels~\cite{joshiHongOuMandelInterferenceFrequency2017,clemmenRamseyInterferenceSingle2016}.

Unfortunately, fiber-based FWM-BS is often accompanied by a broadband spontaneous emission from spontaneous Raman scattering (SpRS) induced by nonlinear interactions of light and localized material vibrations~\cite{clarkHighefficiencyFrequencyConversion2013a,krupaBraggScatteringConversionTelecom2012}.
This can be partly mitigated through cooling~\cite{farsiLowNoiseQuantumFrequency2015a} and partly through restriction to short frequency shifts~\cite{clarkHighefficiencyFrequencyConversion2013a}. 
However, long frequency shifts require a strong pump that is spectrally close to the fragile quantum fields and thus necessitates a more careful analysis of the impact of Raman scattering. 
Appropriate quantum models exist in the context of photon-pair generation~\cite{linPhotonpairGenerationOptical2007, koefoedEffectsNoninstantaneousNonlinear2017} but FWM-BS has only been studied using classical models~\cite{friisEffectsRamanScattering2017} or with a full Lindblad approach~\cite{bonettiSimpleApproachQuantum2019} with the associated computational challenges.

In this work we provide a quantum-mechanical model of FWM-BS in the presence of Raman scattering by exploiting the bilinear nature of the equations of motion. 
This leads to exact solutions for the quantum fields in terms of classical Green function without needing to restrict the Hilbert space. 
Using this model, we provide quantitative guidelines for how the quality of frequency conversion depends on the system configuration. 
We study the photon statistics and spectral properties of converted single photons in the presence of Raman scattering for both continuous wave (CW) pumping and pulsed pumping using a novel numerical split-step scheme.


\section{Theory}
\renewcommand{\q}{\mathrm{q}}

In this section we introduce the evolution equations for the quantum and classical pumps and derive a general expression for the second-order correlation function for the frequency-converted light.

\subsection{Frequency-conversion using Bragg scattering}
\begin{figure}[t]
    \centering
    \includegraphics[width = 0.42\textwidth]{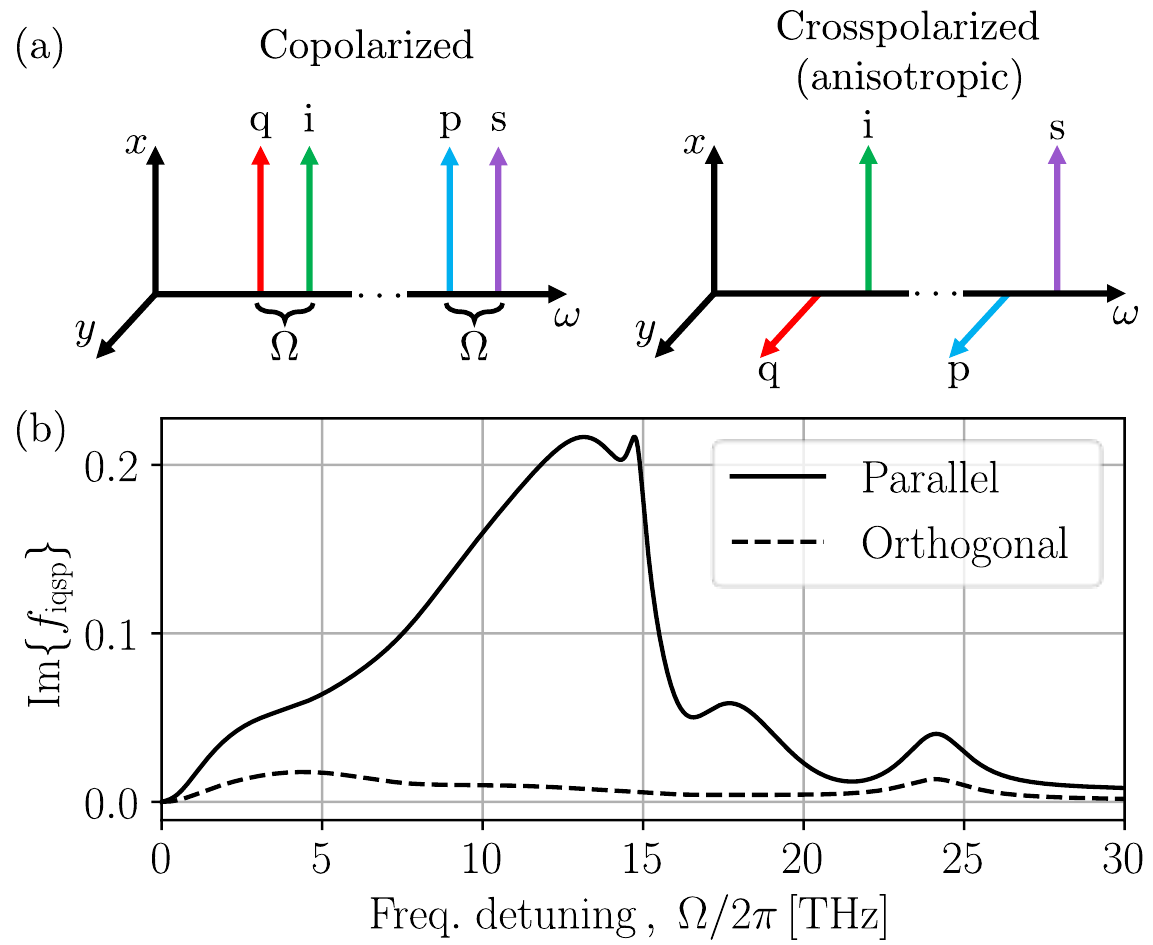}
    \caption{(a) Illustration of two possible polarization configurations of the four fields, idler (i), signal (s), and pumps (p and q). The pumps are separated from the nearest quantum fields by a frequency detuning $\Omega$. (b) The Raman response corresponding to the two polarization configurations for a silica fiber. The data is acquired from refs. \cite{drummondQuantumNoiseOptical2001a,hellwarthThirdorderOpticalSusceptibilities1977}.}
    \label{fig:pol_config}
\end{figure}

FWM-BS is the coherent interaction between four spectrally separated fields as illustrated in \cref{fig:pol_config}(a).
An input signal field (s) is frequency translated into the idler field (i) through a $\chi^{(3)}$-nonlinear interaction mediated by two strong pumps (p and q), under the requirement of energy conservation $\omega_\i - \omega_\s + \omega_\p - \omega_\qi = 0$. For the remainder of our analysis, we assume that the separation $\Omega = \omega_\i - \omega_\q$ is within the Raman bandwidth while the two pumps are separated sufficiently far that we may neglect Raman interactions over this frequency span (that is, there are no phonon modes with a sufficiently large frequency to mediate such interactions).
For linearly polarized fields, the $\chi^{(3)}$-tensor distinguishes between two different polarization configurations of the fields: copolarized and crosspolarized as illustrated in \cref{fig:pol_config}(a).
The copolarized has the largest nonlinear strength, and is therefore usually preferable~\cite{hellwarthThirdorderOpticalSusceptibilities1977}.
However, the Raman response divides the crosspolarized into two: the isotropic and anisotropic configurations~\cite{hellwarthThirdorderOpticalSusceptibilities1977}.
These corresponds to the fields within the Raman bandwidth of each other being copolarized and crosspolarized, respectively.
The copolarized configuration maximizes Raman scattering through the parallel Raman response, whereas the anisotropic minimizes it through the orthogonal Raman response.
The corresponding response functions for a silica fiber is illustrated in \cref{fig:pol_config}(b).
For these reasons we focus on the copolarized and the anisotropic configurations in the following analysis.
It should be noted, that these polarization configurations only hold for linearly polarized fields.
The model presented in this paper is completely general, and is therefore applicable for more complicated polarization modes, such as OAM modes~\cite{rottwittIntermodalRamanAmplification2019}, and spatial mode configurations.
However, for the sake of simplicity, we focus on the linearly polarized modes during the presentation of the model.

\subsection{Field equations}

The full vectorial quantum field equation describing FWM for quantum operators representing the two polarization components have been derived in previous papers~\cite{drummondQuantumTheoryFiberOptics1996a,linPhotonpairGenerationOptical2007}.
In Appendix A, we have deduced the nonlinear coefficients, and response functions when the quantum operators represent arbitrary spatial and polarization modes, and find that the resulting equation of motion is identical to the full vectorial quantum field equation.
To model FWM-BS, we consider the four fields described in the previous section, and only keep the FWM-terms for which $\omega_\p - \omega_\q + \omega_\i -\omega_\s = 0$ is fulfilled.
To simplify the equations we decouple the pump equations from the signal and idler equations by letting the pumps be classical fields which are normalized such that $|A_{\p,\q}|^2$ is the power.
The quantum fields are normalized to have the standard equal-space commutation relations $[\a_n(z,t), \ad_m(z,t')] = \delta_{nm}\delta(t-t')$.
The idler equation is thereby
\begin{align}\label{eq:quantum_field_equation}
    \partial_z \a_\i &= - \beta_{1\i}\partial_t \a_\i - \frac{i}{2}\beta_{2\i}\partial_t^2 \a_\i  \\
    &+ 2i(\gamma_{\i\p} |A_\p|^2 + \gamma_{\i\q}|A_\q|^2) \a_\i + 2\gamma_{\i\p\q\s}A_\p^*A_\q \a_\s\notag\\
    &+ i A_\qi\int\ud t' f_{\i\qi\p\s}(t-t') A_\p^*(z,t')\a_\s(z,t')\notag\\
    &  +i A_\qi\int\ud t' f_{\i\qi\qi\i}(t-t') A_\qi^*(z,t')\a_\i(z,t')+ i\m_{\i\qi} A_\qi, \notag
\end{align}
The indices are contracted such that $\i\p = \i\i\q\q$ and the fields without an argument are evaluated in $(z,t)$.
The instantaneous electronic response is described by $\gamma_{ijkl}$, and the noninstanteous Raman response is represented by $f_{ijkl}$, and each characterize the interaction between the four fields $i$, $j$, $k$ and $l$.
The noise operator $\m_{ij}$ describes the creation/annihilation of a phonon under the interaction between the two fields $i$ and $j$, and thereby represents SpRS.
Both the Raman response function $f_{ijkl}$ and the noise operator $\m_{ij}$ oscillates at frequency $\Omega$, as described in Appendix A.
In addition, the linear dispersion is expanded around the central idler frequency in terms of dispersion parameters $\beta_{1\i}$ and $\beta_{2\i}$. 
In this model, we kept the terms responsible for linear dispersion up to second order, nonlinear phase modulation (NPM) from the pumps, FWM-BS, delayed FWM-BS, delayed NPM, and SpRS. 
We neglected Raman terms that requires interaction with a phonon at the large frequency separation, for example SpRS from pump p to the idler. The signal field equation is identical under the substitution s~$\leftrightarrow$~i and p~$\leftrightarrow$~q. We note that unlike their electronic counterparts, the delayed versions of NPM and FWM-BS do not perfectly conserve the energy of the optical fields, leading to stimulated Raman effects. 

The noise operators are based on the initial thermal phonon distribution and are thus completely determined by their second-order correlation
\begin{subequations}
\begin{align}
    \mean{\md_{ij}(z,\omega)\m_{kl}(z',\omega')} &= \delta(\omega-\omega')\delta(z-z') F_{ijkl}(\omega) \\
    F_{ijkl}(\omega) = \sqrt{2\pi}&\mathrm{Im}[f_{ijkl}(\omega)] n_\mathrm{th} (\omega+\Omega)
\end{align}
\end{subequations}
where $n_\mathrm{th}(\omega) = \left[\exp(\frac{\hbar\omega}{k_\mathrm{B}T}) - 1\right]^{-1}$ is the thermal population number.
Since the field equations are bilinear in the quantum fields, the solution takes the simple input-output form~\cite{ekertRelationshipSemiclassicalQuantummechanical1991,yurkeInputOutputTheory2004} after an interaction of length $z = \ell$
\begin{align}
\label{eq:Green}
    \a_\i(\ell,t) =& \int \ud t' \left [ G_{\i\i}(\ell, t,t') \a_i(0,t) + G_{\i\s}(\ell, t,t')\a_\s(0,t') \right ]\notag\\
    + i&\int \ud t'\int_0^\ell \ud z' \left [A_\q(z',t)G_{\i\i}(\ell-z', t,t')\m_{\i\q}(z',t') \right. \notag \\
    &\quad + \left. A_\p(z',t')G_{\i\s}(\ell-z',t,t')\m_{\s\p}(z',t')\right ],
\end{align}
where $G_{ij}$ are Green functions. 
The first term here represents the frequency-conversion process while the second term represents contamination from SpRS that is also subjected to the frequency-conversion effect after its creation. 

The strong classical pumps are unaffected by the single-photon-level FWM-BS and are thus governed by
\begin{align}\label{eq:pump_equation}
    \partial_z A_\p &= -\beta_{1\p}\partial_t A_\p - \frac{i}{2}\beta_{2\p}\partial_t^2A_\p\\
    &+ \gamma_\p |A_\p|^2 A_\p + 2\gamma_{\p\q}|A_\q|^2 A_\q\notag\\
    &+iA_\p\int\ud t' f_\p(t-t')|A_\p(z,t')|^2\notag\\
    &+iA_\q\int\ud t' f_{\p\q}(t-t')|A_\q(z,t')|^2\notag
\end{align}
The equation for pump $\q$ is found by substituting $\p \leftrightarrow \q$.
By treating the pumps classically, the problem simplifies considerably, since the equations of motions are now bilinear in the signal and idler fields. 
However, as a consequence of the delayed nonlinear response, the optical-field commutators are no longer perfectly preserved because of stimulated Raman scattering from the classical fields, which enters through the Green functions. 
This is most problematic on the Stokes side where the field operators are (clasically) amplified instead of populating higher Fock states. 
On the anti-Stokes side, which is the most experimentally relevant, this behavior is less problematic since the induced decay of the field operator simply lowers their expectation values with respect to the vacuum. 
Calculating the photon flux, the amplification and depletion is captured correctly, however, the second-order correlation function measures the photon statistics, which can therefore not be described properly under a classical amplification.
Since the second-order correlation function is insensitive to loss, the photon statistics of classical depletion is described accurately.

\subsection{Second-order correlation function}

For single-photon applications, the second-order correlation function is an important metric. At the waveguide output $z = \ell$ and at times $t_1, t_2$ it is defined as
\begin{equation}
    g^{(2)}(t_1,t_2) = \frac{\mean{\a_\i^\dagger(\ell,t_1)\a_\i^\dagger(\ell,t_2)\a_\i(\ell,t_2)\a_\i(\ell,t_1)}}{\mean{\a_\i^\dagger(\ell,t_1)\a_\i(\ell,t_1)}\mean{\a_\i^\dagger(\ell,t_2)\a_\i(\ell,t_2)}}.
\end{equation}
At zero delay, a $ g^{(2)}(t,t)$ value of close to 0 indicates strong anti-bunching and thus single-photon statistics at the output at time $t$. For a frequency-converted single photon, a value larger than 0 indicates contamination by Raman noise photons.  
We consider a single-photon input state with a temporal amplitude $\psi(t)$
\begin{equation}
    \ket{\psi} = \int\ud t \, \psi(t) \ad_\s(t)\ket{0}\,,
\end{equation}
for which the second-order correlation function is calculated (see Appendix B for details)
\begin{align}\label{eq:general_g2}
    g^{(2)}(t_1,t_2) =& \frac{1}{I_\mathrm{out}(t_1)I_\mathrm{out}(t_2)}\Big(|\psi_\mathrm{out}(t_1)|^2E(t_2,t_2)\\
    +& |\psi_\mathrm{out}(t_2)|^2E(t_1,t_1) + E(t_1,t_1)E(t_2,t_2)\notag\\
    + 2\mathrm{Re}\{\psi&_\mathrm{out}^*(t_1)\psi_\mathrm{out}(t_2)E(t_1,t_2)\} + |E(t_1,t_2)|^2\Big)\notag
\end{align}
where the expectation values are determined in terms of the Green functions
\begin{subequations}
\begin{align}    
    I_\mathrm{out}(t) =& |\psi_\mathrm{out}(t)|^2 + E(t,t),\\
    \psi_\mathrm{out}(t) =& \int \ud t' G_{\i\s}(\ell,t,t') \psi(t'),\\
    E(t_1,t_2) =& \int_0^\ell\ud z \iint\ud t\ud t' Q^*(z,t_1,t) Q(z,t_2,t')\\
    & \qquad \times F(t-t'),\notag\\
    Q(z,t,t') =& A_\q(z,t)G_{\i\i}(z,t,t') + A_\p(z,t)G_{\i\s}(z,t,t'),
\end{align}
\end{subequations}
We have assumed a balanced Raman response $F = F_{\i\qi\qi\i} = F_{\i\qi\p\s}$, such that the Green functions can be factorized.
Here $F(t)$ is the inverse Fourier transform of $F(\omega)$.
Thus, the problem of calculating $g^{(2)}$ is reduced to the calculation of the Green functions. 

In realistic experiments, we can not access $g^{(2)}(t_1, t_2)$ directly due to the finite temporal resolution of single-photon detectors. Instead, we measure photon detection events ('clicks') within a certain time window $T$. Thus, the experimentally relevant metric is instead~\cite{christProbingMultimodeSqueezing2011}
\begin{equation}\label{eq:g2_click}
    g^{(2)}_\mathrm{click} = \frac{\iint_{-T/2}^{T/2}\ud t_1 \ud t_2 \mean{\a_\i^\dagger(\ell,t_1)\a_\i^\dagger(\ell,t_2)\a_\i(\ell,t_2)\a_\i(\ell,t_1)}}{\left(\int_{-T/2}^{T/2}\ud t \mean{\a_\i^\dagger(\ell,t)\a_\i(\ell,t)}\right)^2}.
\end{equation}
In the following sections, we analyze the second-order photon statistics of frequency-converted single photons in the presence of stimulated and spontaneous Raman scattering for both the case of CW pumps and the experimentally relevant case of pulsed pumps.

\section{Continuous wave pumps}

In this section we analyze the case of continuous-wave (CW) pumps where analytical solutions can be obtained.

\subsection{Signal and idler evolution}
The CW pumps propagate with the pure phase evolution $A_{\p,\qi}(z,t) =\sqrt{P_0}\e^{\i\phi(z)}$.
Defining the Fourier transform as $\a_\i(z,\omega) = \tfrac{1}{\sqrt{2\pi}}\int_{-\infty}^{\infty}\ud t \a_\i(z,t)\e^{\i\omega t}$, we obtain the simple quantum evolution equation for the idler
\begin{equation}
\begin{aligned}
    \partial_z \a_\i(z,\omega) =& \i\kappa_\i(z,\omega)\a_\i(z,\omega)\\
    +& \i g(\omega)\a_\s(z,\omega) + \i \sqrt{P_0} \m_{\i\qi}(z,\omega),
\end{aligned}
\end{equation}
where the coupling functions are given by
\begin{subequations}
    \begin{align}
        \kappa_\i(\omega) &= \beta_\i(\omega) + \sqrt{2\pi}P_0 f_{\i\qi\qi\i}(\omega) + \tfrac{\Delta\beta}{2},\\
        g(\omega) &= 2 P_0 \gamma_{\i\qi\p\s} + \sqrt{2\pi}P_0 f_{\i\qi\p\s}(\omega).
    \end{align}
\end{subequations}
Here, $\kappa_\i$ describes phasemodulation and classical amplification/depletion, induced by the imaginary part of $f_{\i\qi\qi\i}$.
The coupling function $g(\omega)$ describes the gain in the idler field due to the frequency conversion, given by the real part, and stimulated Raman scattering given by the imaginary part.
The signal equation is found under the substitution $\i\leftrightarrow\s$, $\qi\leftrightarrow\p$, and $\Delta\beta \rightarrow -\Delta\beta$.
Due to the pumps being CW the Green functions takes the form $G_{ij}(\ell,t,t') = G_{ij}(\ell,t - t')$, and the solution in the frequency domain is thereby found
\begin{equation}
    \begin{aligned}
        \a_\i(\ell,\omega) = G_{\i\i}(\ell,\omega)\a_\i(0,&\,\omega) + G_{\i\s}(\ell,\omega)\a_\s(0,\omega)\\
        +  \i \sqrt{P_0}\int_0^\ell \ud z \Big[&G_{\i\i}(\ell - z,\omega)\m_{\i\qi}(z,\omega)\\
        + &G_{\i\s}(\ell - z,\omega)\m_{\s\p}(z,\omega)\Big]
    \end{aligned}
\end{equation}
where the Green functions are given by
\begin{subequations}
    \begin{align}
    G_{\i\i}(z,\omega) &=\frac{\e^{\tfrac{i}{2}[\kappa_\s(\omega) + \kappa_\i(\omega)]z}}{k(\omega)}\Big\{ k(\omega)\cos(k(\omega)z)\notag\\
    &- \tfrac{i}{2}\Delta\beta(\omega)\sin(k(\omega)z)\Big\} \\
    G_{\i\s}(z,\omega) &= i\frac{g(\omega)}{k(\omega)}\e^{\tfrac{i}{2}[\kappa_\s(\omega) + \kappa_\i(\omega)]z}\sin(k(\omega)z)
    \end{align}
\end{subequations}
and we have defined the effective coupling function due to phase mismatch, field walk-off, and higher-order dispersion, as
\begin{equation}
    k(\omega) = \tfrac{1}{2}\sqrt{4g(\omega)^2 + \Delta\beta(\omega)^2}\,.
\end{equation}
From the explicit expressions of the Green functions the classical amplification/depletion of the quantum fields are seen to enter through the imaginary part of $\kappa_{\s,\i}$ and $k$.

\subsubsection{Spectral filtering}
In practice the idler is filtered spectrally to reduce the amount of spontaneously emitted Raman photons.
The filtered quantum field can be written as
\begin{equation}\label{eq:output_filter}
    \a_\i(\ell,t) = \frac{1}{\sqrt{2\pi}}\int_{-\infty}^\infty\ud\omega H(\omega) \a_\i(\ell,\omega) \e^{- i\omega t}
\end{equation}
where $H(\omega)$ is the spectral filter transmission function, which is centered around the center frequency of the idler.
The output temporal distribution of the photon state is thereby
\begin{equation}\label{eq:photon_expec_CW}
    \psi_\mathrm{out}(t) = \frac{1}{2\pi}\int\ud t' \int\ud\omega H(\omega) G_{\i\s}(\ell,\omega) \psi(t') \e^{-i\omega(t-t')}\,.
\end{equation}
The expectation value of the phonon operators is similarly found to be
\begin{align}\label{eq:phonon_expec_CW}
    E(t_1,t_2) =& \frac{P_0}{\sqrt{2\pi}}\int_0^\ell \ud z \int\ud \omega |H(\omega)|^2\\
    \times &|G_{\i\i}(z,\omega) + G_{\i\s}(z,\omega)|^2 F(\omega) \e^{-i\omega (t_1 - t_2)}\,.\notag
\end{align}
In the following, we consider a rectangular spectral filter with spectral width $\Delta\omega$.

\subsubsection{Linearly polarized fields}
The results presented in the previous sections are completely general as they describe any field mode or polarization.
In the following examples of the model, we restrict our attention to the experimentally relevant situation of linearly polarized fields.
Additionally, we parameterise the response functions through the Raman fraction $\fR$, which describes the fraction of the nonlinear response that is attributed to Raman scattering.
The electronic and Raman response functions thereby take the form~\cite{linPhotonpairGenerationOptical2007}
\begin{subequations}
\begin{align}
    \gamma_{ijkl} &= \frac{1}{2}\gamma(1-\fR)(\delta_{ij}\delta_{kl} + \delta_{ik}\delta_{jl} + \delta_{il}\delta_{jk}),\\
    f_{ijkl}(t) &= \gamma \fR\e^{i\Omega t} \bigg(h_a(t)\delta_{ij}\delta_{kl}\\
    & \qquad + \frac{1}{2}h_b(t)\left[\delta_{ik}\delta_{jl} + \delta_{il}\delta_{jk}\right]\bigg),\notag
\end{align}
\end{subequations}
where the indices on the Kroenecker deltas refer to the polarization of the corresponding mode, and $\gamma$ depends on the spatial mode overlap of the field profiles, as described in Appendix A.
In the remanining, we assume a single mode fiber, such that $\gamma$ is the same for NPM and Bragg scattering.
The sum of the isotropic-orthogonal $h_a$ and anisotropic-orthogonal $h_b$ Raman response gives the parallel response $h$, which can be estimated as a sum of harmonic oscillators
\begin{equation}
    h(t) = \Theta(t) \sum_i  F_i d_i \sin(\omega_i t)\e^{-d_i t},
\end{equation}
where $\Theta(t)$ is the Heaviside step-function. 
Using 11 oscillators provides a good approximation for the silica Raman response function~\cite{drummondQuantumNoiseOptical2001a}.
From the parallel response function we extract the anisotropic response function using Ref.~\cite{hellwarthThirdorderOpticalSusceptibilities1977}.

\subsection{Photon flux}
\begin{figure}[t]
    \centering
    \includegraphics[width = 0.4\textwidth]{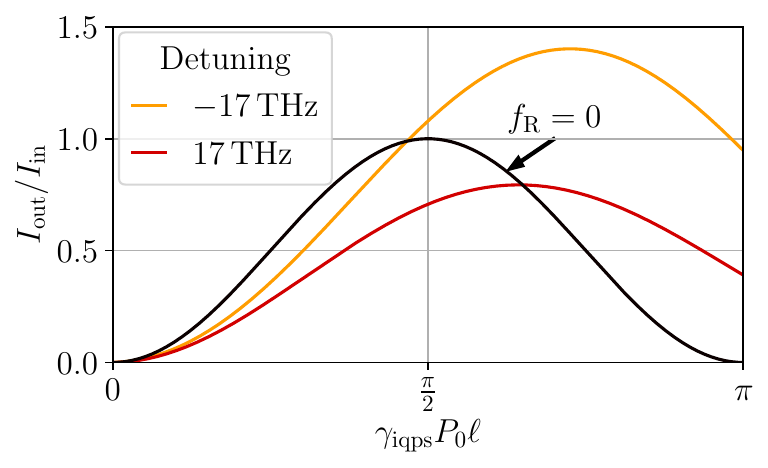}
    \caption{The relative output flux, as a function of the dimensionless nonlinear interaction strength $\gamma P_0 \ell$ in a co-polarized configuration. The coloured lines include Raman scattering, and are illustrated for $\fR = 0.18$, detuning $\Omega/2\pi$, and $\Delta\omega/I_\mathrm{in} = 1$.}
    \label{fig:flux_CW}
\end{figure}
To illustrate how Raman scattering impacts the general frequency conversion process, we start by investigating the frequency-converted photon flux $I_\mathrm{out}$.
We consider a single-photon with a spectral width much smaller than the variations of the Raman response, being $\frac{1}{\tau_\p}\ll\mathrm{min}\{d_i\}$.
The filter width is decreased correspondingly, such that $\Delta\omega \ll \mathrm{min}\{d_i\}$.
For the remainder of this section we assume perfect phase matching $\Delta\beta(\omega) = 0$ corresponding to $k(\omega) = g(\omega)$.
To first order in the filter width, the output flux thereby becomes
\begin{align}
    I_\mathrm{out} = &|G_{\i\s}(\ell,\omega = 0)|^2 I_\mathrm{in}\\
    + &\frac{\Delta\omega}{4\sqrt{2\pi}} \Big(1- \exp[-8\pi \fR R(0)\gamma P_0\ell]\Big) n_\mathrm{th}(\Omega)\notag
\end{align}
where $\mathrm{Im}\{f(\omega)\} = \sqrt{2\pi}\gamma \fR R(\omega)$, and the indices have been suppressed for compactness.
In \cref{fig:flux_CW} the relative output flux is shown as a function of the dimensionless nonlinear interaction strength $\gamma_\mathrm{iqps} P_0 \ell$, in a Stokes ($\Omega/2\pi = -17\,\mathrm{THz}$), and an anti-Stokes ($\Omega/2\pi = 17\,\mathrm{THz}$) configuration, with $\fR = 0.18$.
These are compared to the relative output flux in the absence of Raman scattering $\fR = 0$.

The optimal interaction strength in absence of Raman scattering is given by $\gamma_\mathrm{iqps} P_0\ell = \tfrac{\pi}{2}$, as can be seen directly from the Green function $G_{\i\s}$.
As Raman scattering is introduced, the relative output flux significantly depends on the frequency detuning.
Due to stimulated Raman scattering, the Stokes configuration experiences gain, whereas the anti-Stokes configuration experiences depletion.
The impact of stimulated Raman scattering can be seen more easily through the explicit expression of the Green function
\begin{equation}
    |G_{\i\s}|^2 = \frac{\e^{-2\mathrm{Im}\{\kappa\}\ell}}{2}\left(\cosh(2\mathrm{Im}\{g\}\ell) - \cos(2\mathrm{Re}\{g\}\ell)\right).
\end{equation}
The imaginary part of the cross-field coupling function $g$ induces a gain through Raman-induced Bragg scattering, effectively constituting a four-field stimulated Raman scattering effect.
However, the imaginary part of the self-field coupling function $\kappa$, introduces gain or absorption, depending on the sign of $\Omega$, and is the well-known stimulated Raman scattering effect.
Evidently stimulated absorption requires an anti-Stokes configuration, corresponding to $\mathrm{Im}\{\kappa\}>0$, and $\mathrm{Im}\{g-\kappa\}< 0$.
In the examples presented here, we consider a balanced Raman response $\mathrm{Im}\{g-\kappa\}= 0$.
To see the impact of spontaneous Raman scattering it is necessary to investigate the second-order correlation function.


\subsection{Second-order correlation}
\begin{figure}[b]
    \centering
    \includegraphics[width = 0.35\textwidth]{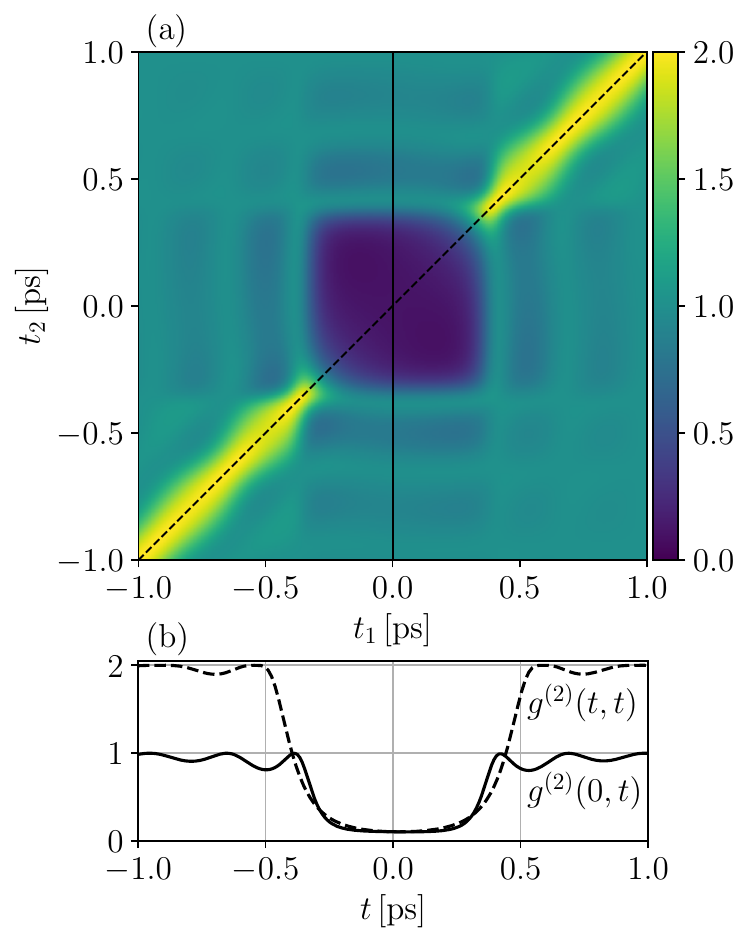}
    \caption{(a) The second-order correlation function $g^{(2)}(t_1,t_2)$ illustrated for $\tau_p = 0.1\,\mathrm{ps}$, $\fR = 0.18$, $\Omega/2\pi = -17\,\mathrm{THz}$, and a temperature of 300\,K. (b) The cross-sections of $g^{(2)}$ given by the solid and dashed lines in (a).}
    \label{fig:inst_g2_CW}
\end{figure}
Having determined the output probability amplitude $\psi_\mathrm{out}$, and the phonon expectation value \cref{eq:phonon_expec_CW}, the second-order correlation function is directly determined by \cref{eq:general_g2}.
We choose to investigate an input state with a Gaussian temporal amplitude, given by
\begin{equation}
    \psi(t) = \frac{1}{\sqrt{\tau_p}(2\pi)^{\tfrac{1}{4}}}\e^{-\tfrac{t^2}{4\tau_p^2}}\,,
\end{equation}
where $\tau_p$ is the standard deviation of the probability distribution $|\psi|^2$.
For the present analysis we focus on photon pulses with a duration on the order of the Raman response $\tau_\p = 0.1$\,ps.
The filter width is chosen corresponding to the pulse duration $\Delta\omega = \tfrac{2}{\tau_p}$, so that the photon with high probability passes through the filter.

In \cref{fig:inst_g2_CW}(a) the second-order correlation function given by \cref{eq:general_g2} is shown.
Along the dashed line in \cref{fig:inst_g2_CW}(b), the $g^{(2)}$-value at zero delay time $g^{(2)}(t,t)$ is found.
From this, the idler-field is seen to exhibit single-photon statistics when the single-photon is present $|t_1| = |t_2| < \tau_p$.
Outside the time window of the single photon the idler field exhibits thermal statistics, which indicates the presence of spontaneous Raman photons, which are inherently thermal.
The solid line in \cref{fig:inst_g2_CW}(b) corresponds to the $g^{(2)}$-value as a function of delay time, measured exactly at the single-photon peak.

\begin{figure}[b]
    \centering
    \includegraphics[width = 0.35\textwidth]{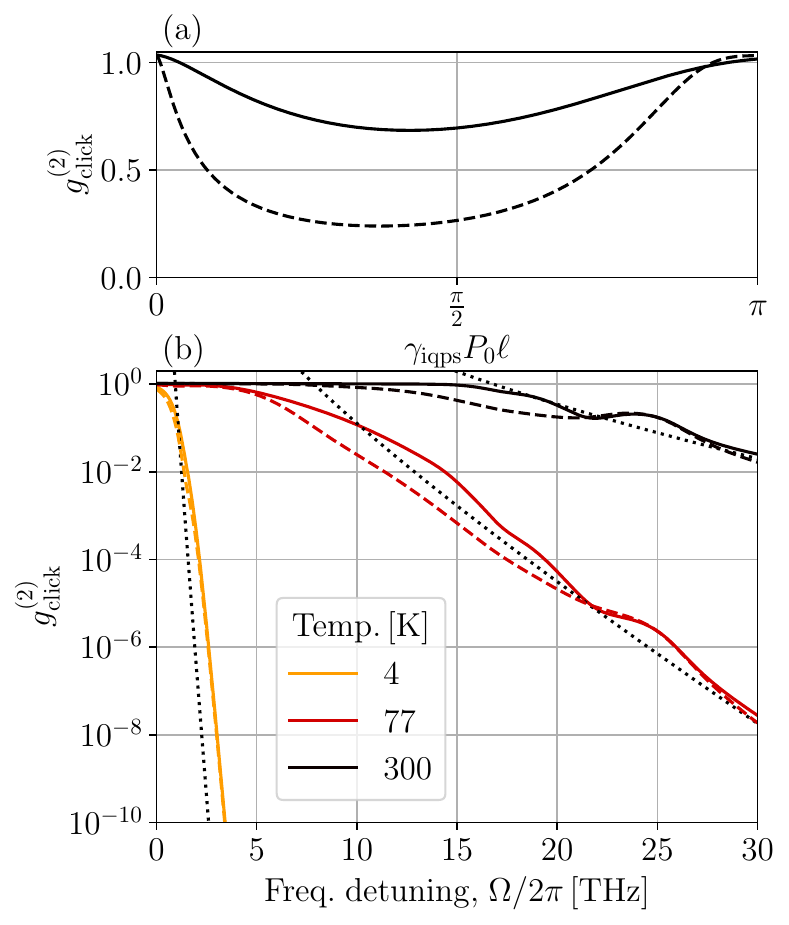}
    \caption{(a) The time-resolved second-order correlation value $g_\mathrm{click}^{(2)}$ as function of nonlinear interaction strength, with $\gamma P_0 = 1\,(\mathrm{km\,W})^{-1}$, $\fR = 0.18$, a fiber temperature of 300\,K, and a frequency detuning of $\Omega/2\pi = -17\,\mathrm{THz}$. (b) $g_\mathrm{click}^{(2)}$ as a function of frequency detuning. This is illustrated for the co-polarized (solid line) and the anisotropic (dashed line) polarization configuration, for three different fiber temperatures. The black dotted lines corresponds to the approximated $g_\mathrm{click}^{(2)}$ value \cref{eq:g2_approx} for the corresponding temperatures.}
    \label{fig:integrated_g2_CW}
\end{figure}

We now consider the experimentally relevant time-resolved $g_\mathrm{click}^{(2)}$-value, with a time-window of 10\,ps, on the order of realistic detector resolutions~\cite{esmaeilzadehEfficientSinglePhotonDetection2020}.
In \cref{fig:integrated_g2_CW}(a) $g_\mathrm{click}^{(2)}$ is shown as a function of the nonlinear interaction strength, with a varying interaction length, for a frequency detuning of $\Omega/2\pi = 17\,\mathrm{THz}$, and a fiber temperature of 300\,K.
Initially, the field is dominated by the spontaneous Raman photons, which is characterized by a value of $g_\mathrm{click}^{(2)}\sim 1$.
Here, the thermal statistics are not apparent since the thermal correlations are much shorter than the detector window due to the broadband nature of the Raman spectrum.
As the interaction length increases, $g_\mathrm{click}^{(2)}$ decreases, indicating the conversion of the single photon.
The anisotropic polarization configuration yields a significantly lower $g_\mathrm{click}^{(2)}$-value, compared to the co-polarized configuration, even though the appropriate interaction length is a factor of three larger.
This is primarily caused by the orthogonal Raman response being approximately a factor 10 smaller.
The optimal $g_\mathrm{click}^{(2)}$-value is not obtained at the optimal interaction length due to the linear increase in spontaneously generated Raman photons.

In \cref{fig:integrated_g2_CW}(b) $g_\mathrm{click}^{(2)}$ is shown as a function of frequency detuning at the optimal interaction length, for three different temperatures, and the two polarization configurations.
The Raman response is clearly present for the two larger temperatures, which again shows the significance of the anisotropic response.
For the low temperature $4\,\mathrm{K}$, the phonon occupation is almost non-existent, thereby suppressing the amount of spontaneous Raman scattering.

In the example presented here, the single-photon duration is on the order of the Raman response $\sim 0.1$\,ps.
However, single-photons produced by quantum dots are usually on the order of nanoseconds, permitting a filter-width on the order of GHz~\cite{lodahlInterfacingSinglePhotons2015}.
We therefore consider the regime in which the spectral filter width $\Delta\omega$ is much smaller than the variations of the Raman response.
Additionally, we assume that the photon is completely captured both spectrally and temporally, such that $T\Delta\omega \gg 1$.
Finally, we assume that the effect of the spontaneous Raman photons are smaller than the frequency conversion of the single-photon corresponding to $|T\Delta\omega \fR \gamma P_0\ell R(0)| \ll 1$.
Under these three assumptions, we obtain the approximate result
\begin{equation}\label{eq:g2_approx}
    g_\mathrm{click}^{(2)} \approx \left(T\Delta\omega + 2\pi\right) \sqrt{2\pi} R(0)n_\th(\Omega)\fR\gamma P_0\ell.
\end{equation}
Evidently $g_\mathrm{click}^{(2)}$ is proportional to $R\,n_\th$, which is the spontaneous Raman spectrum.
In the regime $|\Omega| \rightarrow\infty$ the Raman response can be expanded in a Laurent series such that $R(0) \approx \frac{1}{\Omega^3}\sum_i F_i\omega_i d_i^2$, which is used in \cref{fig:integrated_g2_CW}(b) to illustrate the approximated $g_\mathrm{click}^{(2)}$-value \cref{eq:g2_approx}, as the black dotted lines.
Even though $\Delta\omega = 20\,\mathrm{THz}$ is much larger than the variations of the Raman response in \cref{fig:integrated_g2_CW}(b), the approximated $g_\mathrm{click}^{(2)}$-value estimates the order of magnitude for the exact value, for the temperatures 300\,K and 77\,K.
Although, for 4\,K, the approximation error is emphasized due to $n_\th$ being small.

The analytical results show, that it is possible to achieve a second-order correlation value on the single-photon level within the Raman bandwidth by properly engineering the fields, through frequency detuning, polarization configuration, and fiber temperature.
However, using pulsed pumps, unlocks additional degrees of freedom, and we consider this case in the following section.

\section{Pulsed pumps}

In this section, we present a general evolution equation for the Green functions introduced in \cref{eq:Green}. In contrast to previous approaches, which involve propagating modes of a Schmidt decomposition without Raman scattering~\cite{reddySortingPhotonWave2015,mejlingEffectsNonlinearPhase2012}, we directly solve for the Greens function.
We obtain the governing equations by inserting the proposed solution \cref{eq:Green} into \cref{eq:quantum_field_equation}.
The resulting Green function equation is expressed as a vector equation on the form
\begin{equation}\label{eq:GF_equation}
    (\partial_z + \tens{\beta}_1\partial_t + \tfrac
    {i}{2}\tens{\beta}_2 \partial_t^2) \mathbf{G}(z,t,t') = i\int \ud \tau \tens{K}(z,t,\tau)\mathbf{G}(z,\tau,t')
\end{equation}
where the Green function vector, and dispersion coefficient matricies are given by
\begin{equation}
    \mathbf{G}(z,t,t') =
    \begin{pmatrix}
        G_{\i\i}(z,t,t')\\
        G_{\s\i}(z,t,t')
    \end{pmatrix}\,,
    \qquad\tens{\beta}_n =
    \begin{pmatrix}
        \beta_{n\i} & 0\\
        0 & \beta_{n\s}
    \end{pmatrix}.
\end{equation}
The coupling matrix $\tens{K}$ has an instantaneous and noninstantaneous part, representing the electronic, $\tens{K}_\mathrm{E}$ and Raman $\tens{K}_\mathrm{R}$ response, respectively.
These are given by
\begin{subequations}
\begin{align}
    \tens{K}(z,t,t') &= \tens{K}_\mathrm{E}(z,t)\delta(t-t') + \tens{K}_\mathrm{R}(z,t,t'),\\
    \tens{K}_\mathrm{E}(z,t) &= 
    \begin{pmatrix}
        K_{\i\qi\qi\i}^\mathrm{E}(z,t) & K_{\i\qi\p\s}^\mathrm{E}(z,t)\e^{i\Delta\beta z}\\
        K_{\s\p\qi\i}^\mathrm{E}(z,t)\e^{-i\Delta\beta z} & K_{\s\p\p\s}^\mathrm{E}(z,t)
    \end{pmatrix},\\
    \tens{K}_\mathrm{R}(z,t,t') &=
    \begin{pmatrix}
        K_{\i\qi\qi\i}^\mathrm{R}(z,t,t') & K_{\i\qi\p\s}^\mathrm{R}(z,t,t')\e^{i\Delta\beta z}\\
        K_{\s\p\qi\i}^\mathrm{R}(z,t,t')\e^{-i\Delta\beta z} & K_{\s\p\p\s}^\mathrm{R}(z,t,t')
    \end{pmatrix},
\end{align}
\end{subequations}
where the matrix elements are defined as
\begin{subequations}
\begin{align}
    K_{ijkl}^\mathrm{E}(z,t) &= 2\gamma_{ijkl}A_k^*(z,t)A_j(z,t),\\
    K_{ijkl}^\mathrm{R}(z,t,t') &= f_{ijkl}(t-t')A_k^*(z,t')A_j(z,t).
\end{align}
\end{subequations}
For the present analysis, the equations for $G_{\i\i}$ and $G_{\s\i}$ are sufficient since $G_\mathrm{is} = G_\mathrm{si}^\dagger$.
However, the equation for the two remaining Green functions is found under the substitution $\i \leftrightarrow\s$ and $\qi \leftrightarrow\p$.
From \cref{eq:Green} the initial condition to the Green function equation is inferred, and is given by
\begin{equation}
    \mathbf{G}(z=0,t,t') = 
    \begin{pmatrix}
        1\\
        0
    \end{pmatrix}
    \delta(t-t').
\end{equation}
The Green functions depend on the evolving pump fields.
Therefore, to solve the Green function equation, it is necessary to simultaneously solve \cref{eq:pump_equation}, which is carried out in the following using the conventional split-step Fourier method~\cite{agrawalWAVEPROPAGATIONOPTICAL1995}.
Before giving an example of the solutions we outline how the split-step Fourier method is applied on the Green function equation.

\subsection{Split-step scheme for Green functions}
In this section we describe a completely general split-step scheme for solving the Green function equation.
The spatial evolution of the Green functions can be expressed as
\begin{equation}\label{eq:Green_function_operator_equation}
    \partial_z\mathbf{G}_z = \left(D + E_z + R_z\right)\mathbf{G}_z,
\end{equation}
where $D$, $E_z$ and $R_z$ are operators, representing dispersive effects, electronic and Raman-induced NPM and Bragg scattering at position $z$, respectively.
The operators only act on the first time argument, as described by \cref{eq:GF_equation}.
Solving the equation over an increment $h$, and invoking the trapezoidal rule to approximate the integral in the exponential, the formal solution can be expressed as
\begin{equation}
    \mathbf{G}_{z+h} = \exp\Big[Dh + \tilde{E}_z h + \tilde{R}_z h\Big]\mathbf{G}_z + \mathcal{O}(h^3),
\end{equation}
where $\tilde{O}_z = \tfrac{1}{2}(O_{z+h} + O_z)$.
From here, the Baker-Hausdorf formula~\cite{agrawalWAVEPROPAGATIONOPTICAL1995} can be applied twice, thereby separating the evolution of each of the physical effects, giving rise to a symmetrical split-step scheme with three operators, and a local step error of $\mathcal{O}(h^3)$.
\begin{figure}[t]
    \centering
    \includegraphics[width = 0.45\textwidth]{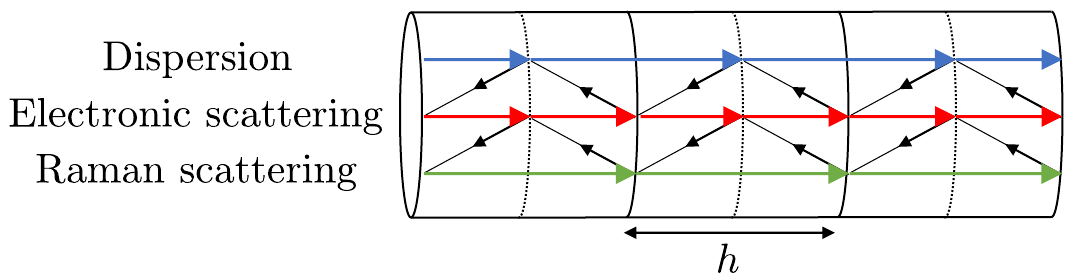}
    \caption{Illustration of the incremental evolution of the three operators in the split-step scheme, applied on three iterated steps. Starting from the first $\tfrac{h}{2}$-dispersion step, the black arrows illustrate the chronological order of the scheme.}
    \label{fig:split_step}
\end{figure}
In \cref{fig:split_step} the symmetrical split-step scheme is illustrated for a single step, and the consecutive application of the scheme.
Each step is propagated by determining the impact of each operator, which is found by solving the equation for each physical effect.
The dispersive and electronic effects has the exact solutions
\begin{subequations}
\begin{align}
    \left[\e^{D_z h}\mathbf{G}_z\right](\omega,t') &= \exp\left[i(\tens{\beta}_1\omega + \tfrac{1}{2}\tens{\beta}_2\omega^2)h\right]\mathbf{G}_z(\omega,t'),\\
    \left[\e^{\tilde{E}_z h}\mathbf{G}_z\right](t,t') &= 
    \begin{pmatrix}
        \mathcal{G}_{\i\i}(z,t) & \mathcal{G}(z,t)\\
        -\mathcal{G}^*(z,t) & \mathcal{G}_{\s\i}(z,t)
    \end{pmatrix}
    \mathbf{G}_z(t,t'),
\end{align}
\end{subequations}
where the frequency argument in the dispersive step implies a Fourier transform.
We have defined the matrix elements of the electronic step as
\begin{subequations}
\begin{align}
    \mathcal{G}_{\i\i}(z,t) &= \frac{\e^{ik_1(z,t) - \tfrac{i}{2} \Delta\beta h}}{k_2(z,t)}\Big\{k_2(z,t) \cos(k_2(z,t) h)\\
    + \tfrac{i}{2}&[\tilde{K}_{\i\qi\qi\i}^\e(z,t) - \tilde{K}_{\s\p\p\s}^\e(z,t) + \Delta\beta]\sin(k_2(z,t) h)\Big\},\notag\\
    \mathcal{G}(z,t) &= i\frac{\tilde{K}_{\i\qi\p\s}^\e(z,t)}{k_2(z,t)}\e^{ik_1(z,t) - \tfrac{i}{2} \Delta\beta h} \sin(k_2(z,t) h),\\
    k_1(z,t) &= \frac{1}{2}[\tilde{K}_{\i\qi\qi\i}^\e(z,t) + \tilde{K}_{\s\p\p\s}^\e(z,t)],\\
    k_2(z,t) &= \frac{1}{2}\Big\{4|\tilde{K}_{\i\qi\p\s}^\e(z,t)|^2\\
    &+ [\tilde{K}_{\i\qi\qi\i}^\e(z,t) - \tilde{K}_{\s\p\p\s}^\e(z,t) + \Delta\beta]^2\Big\}^{1/2},\notag
\end{align}
\end{subequations}
where $\mathcal{G}_{\s\i}$ is found from $\mathcal{G}_{\i\i}$ by substituting $\i \leftrightarrow \s$, and $\Delta\beta \rightarrow - \Delta\beta$.
The tilde denotes the average over the $z$ and $z+h$ points.
Notably, the solution to the pure electronic scattering has the same form as the solution in the CW regime.

Finally, we solve the Raman step.
The equation governing this step is an integro-differential equation, which has a form that does not have a general analytical solution.
The equation can be solved using numerical ODE solvers, however, here we propose a simple iterative approach.
We start by integrating $\hat{R}_z\mathbf{G}_z$ over the increment $h$, and invoking the trapezoidal rule to approximate the $z$-integral
\begin{subequations}
\begin{align}
    \left[\e^{\tilde{R}_z h}\mathbf{G}_z\right](t,t') &= \mathbf{G}_z(t,t') + \Delta\mathbf{G}_z(t,t')\,,\\
    \Delta\mathbf{G}_z(t,t') &= i\frac{h}{2}\int \ud \tau \Big[\tens{K}_\mathrm{R}(z,t,\tau)\mathbf{G}_z(\tau,t')\\
    &+ \tens{K}_\mathrm{R}(z+h,t,\tau)\mathbf{G}_{z+h}(\tau,t')\Big]\notag .
\end{align}
\end{subequations}
We determine $\Delta\mathbf{G}_z$ recursively by first assuming $\mathbf{G}_{z+h} = \mathbf{G}_{z}$.
After three iterations the error is $\mathcal{O}(h^3)$, which is consistent with the error of the symmetric split-step scheme.

\subsection{Numerical results}
In this section, we apply the scheme developed in the previous section to the simple case of an optical fiber with the fields placed symmetrically around the zero-dispersion line.
The fields thus copropagate pairwise with $\beta_{1\s} = \beta_{1\qi}$ and $\beta_{1\i} = \beta_{1\p}$, and walk-off $\beta_{1\s} - \beta_{1\i} = \frac{\Delta n_\mathrm{eff}}{c}$, where $c$ is the speed of light and we choose $\Delta n_\mathrm{eff} = 10^{-3}$.
For simplicity, the input pumps are chosen to be identical with a Gaussian pulse form
\begin{equation}
    A_{\qi,\p}(0,t) = \sqrt{P}\exp\left[\frac{-(t\pm\tfrac{\Delta t}{2})^2}{4\tau_\p^2}\right],
\end{equation}
where $P$ is the peak power, $\tau_\p$ is the pulse duration, and $\Delta t$ is the initial separation between the pumps.
The fiber length is chosen so that the pump fields collide at the fiber midpoint, corresponding to $\ell = \tfrac{2\Delta t c}{\Delta n_\mathrm{eff}}$, and to ensure complete pump-pump collision, we choose $\Delta t = 6 \tau_\p$.
To highlight the phonon-induced temporal dynamics, we only consider pulse durations on the order of the Raman response $\tau_\p = 0.1\,\mathrm{ps}$.

\begin{figure}[t]
    \centering
    \includegraphics[width = 0.48\textwidth]{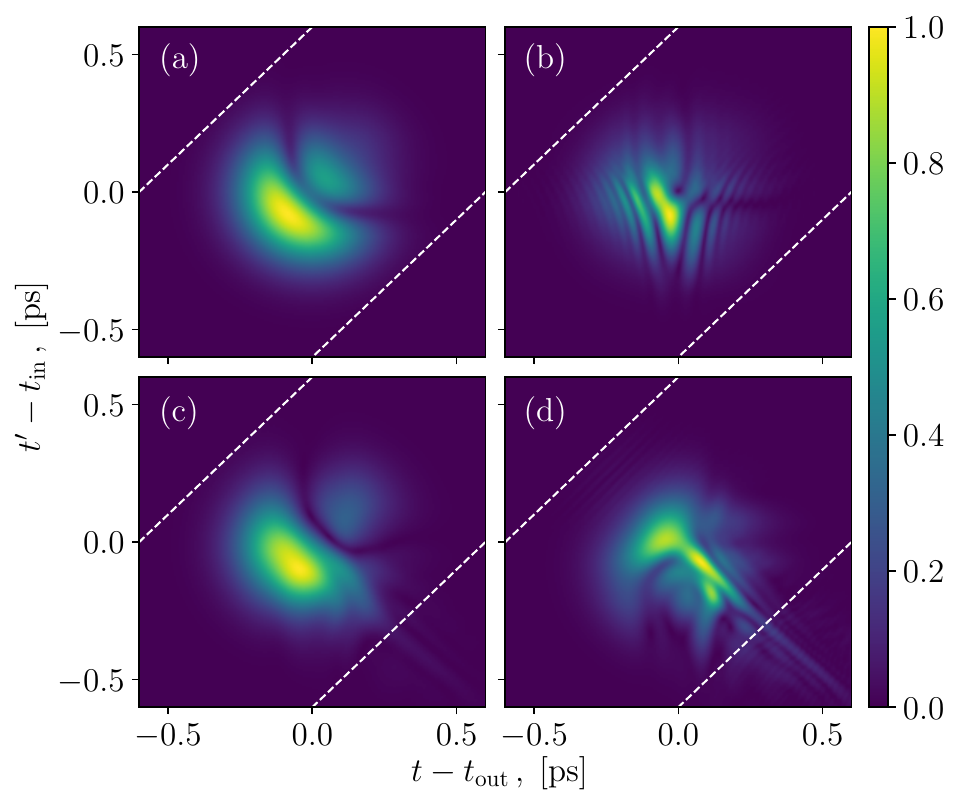}
    \caption{The cross-Green functions $|G_{\i\s}|$ illustrated for 4 choices of parameters each normalized to their respective maximum. (a) Pure electronic response $\fR = 0$. (b) Pure electronic response $\fR = 0$, and second order dispersion $\beta_{2\s,\p} = -\beta_{2\i,\qi} = 0.01\,\mathrm{\tfrac{ps}{m}}$. (c) Raman fraction corresponding to a silica fiber $\fR = 0.18$. (d) Pure Raman response $\fR = 1$.}
    \label{fig:GF_plot}
\end{figure}

In \cref{fig:GF_plot} the cross-Green function is illustrated for $\gamma P \ell = 4\pi$ which ensures full conversion for a suitably chosen input shape.
Lines of causality, illustrated by the white dashed lines, inherently bound the cross-Green function for a pure electronic response due to the instantaneous nature of the electrons.
The lines are given by $t_{\i,\s} = t' + \beta_{1\i,\s}\ell$, with $t_\i$ being the upper line, and $t_\s$ the lower line, corresponding to the earliest and latest times at which a frequency converted photon can arrive.
Fig. \ref{fig:GF_plot}(a) shows $|G_{\i\s}|$ for a pure electronic response.
It is characterized by the shape of the pump, and a ripple.
The same behaviour has been observed in the analytical solutions, which have previously been determined for a pure electronic response, where the ripple enters through the zeroth-order Bessel function of the first kind~\cite{mckinstrieQuantumstatepreservingOpticalFrequency2012}.
Introducing second-order dispersion in \cref{fig:GF_plot}(b), a complicated overlying pattern appears due to interference between the possible paths of conversion.
In \cref{fig:GF_plot}(c-d) Raman scattering is included with $\fR = 0.18$ and $\fR = 1$, respectively.
For $\fR = 0.18$ the Green function is mildly distorted while for $\fR = 1$ the Green function is significantly altered, and it extends beyond the slow (electronic) causality line due to a delay induced by the noninstantaneous Raman response.

\begin{figure}[t]
    \centering
    \includegraphics[width = 0.45\textwidth]{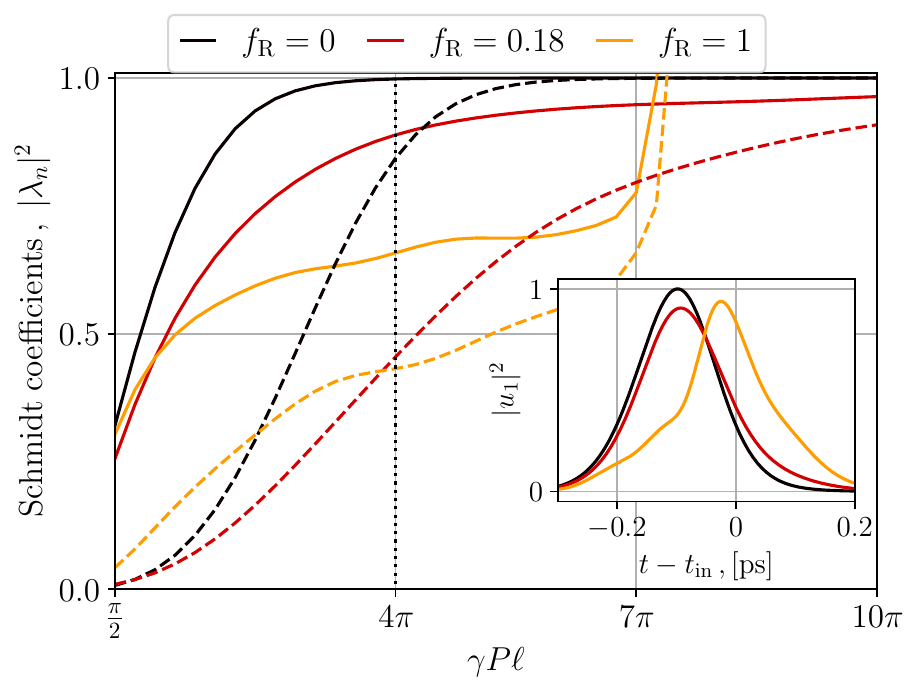}
    \caption{The first (solid line) and second (dashed line) Schmidt mode coefficients as a function nonlinear interaction strength $\gamma P\ell$ found by varying the pump peak power $P$. These are shown for three choices of Raman fraction: pure electronic response $\fR = 0$, silica fiber $\fR = 0.18$, and pure Raman response $\fR = 1$. The inset shows the first-order input Schmidt mode for each of the three responses, evaluated at $\gamma P \ell = 4\pi$.}
    \label{fig:schmidt_mode}
\end{figure}

To investigate the temporal dynamics, it is convenient to decompose the Green functions into Schmidt modes, which explicitly reveals the temporal shapes that are converted.
The Schmidt decomposition of the Green function has the form
\begin{equation}
    G_{\i\s}(t,t') = \sum_n \lambda_n v_n(t) u_n(t'),
\end{equation}
where $\lambda_n$ is the Schmidt coefficient describing the weight of each Schmidt mode, $v_n$ is the output mode, and $u_n$ is the input mode. 
If the $n$th input Schmidt mode is chosen as the input, it is converted to the corresponding $n$th output Schmidt mode with conversion efficiency $|\lambda_n|^2$.
In \cref{fig:schmidt_mode} the Schmidt coefficients for the two first Schmidt modes are shown as a function of the nonlinear interaction strength $\gamma P\ell$, for three choices of $\fR$.
For the pure electronic response $\fR = 0$, the first-order Schmidt coefficient dominates initially, indicating a separable Green function in the low-interaction regime.
As $\gamma P\ell$ increases, the second-order Schmidt coefficient becomes significant, which is seen as the emergence of a ripple in the Green function \cref{fig:GF_plot}(a).
For sufficiently large nonlinear interaction strength both Schmidt modes are perfectly converted.
The same behavior was observed in analytical models~\cite{mckinstrieQuantumstatepreservingOpticalFrequency2012}.

Introducing Raman scattering, $\fR = 0.18$, the Schmidt coefficients decrease primarily due to absorption from stimulated Raman scattering.
For $\fR =1$, the absorption is more significant, and notably the second-order Schmidt coefficient increases faster initially, indicating that Raman scattering deteriorates the separability of the Green function.
Around $\gamma P\ell = 7\pi$ the gain induced by phonon-mediated FWM and absorption from stimulated Raman scattering no longer balances each other due to phase-modulation effects in the pumps, which gives rise to an exponential increase of the Schmidt coefficients.

The inset in \cref{fig:schmidt_mode} shows the first-order input Schmidt modes for the three choices of $\fR$, evaluated for $\gamma P \ell = 4\pi$.
The input Schmidt modes including Raman scattering acquires a slight temporal broadening due to the delayed Raman response, which thereby converts more efficiently before the point of collision compared to the pure electronic response.

\begin{figure}[t]
    \centering
    \includegraphics[width = 0.45\textwidth]{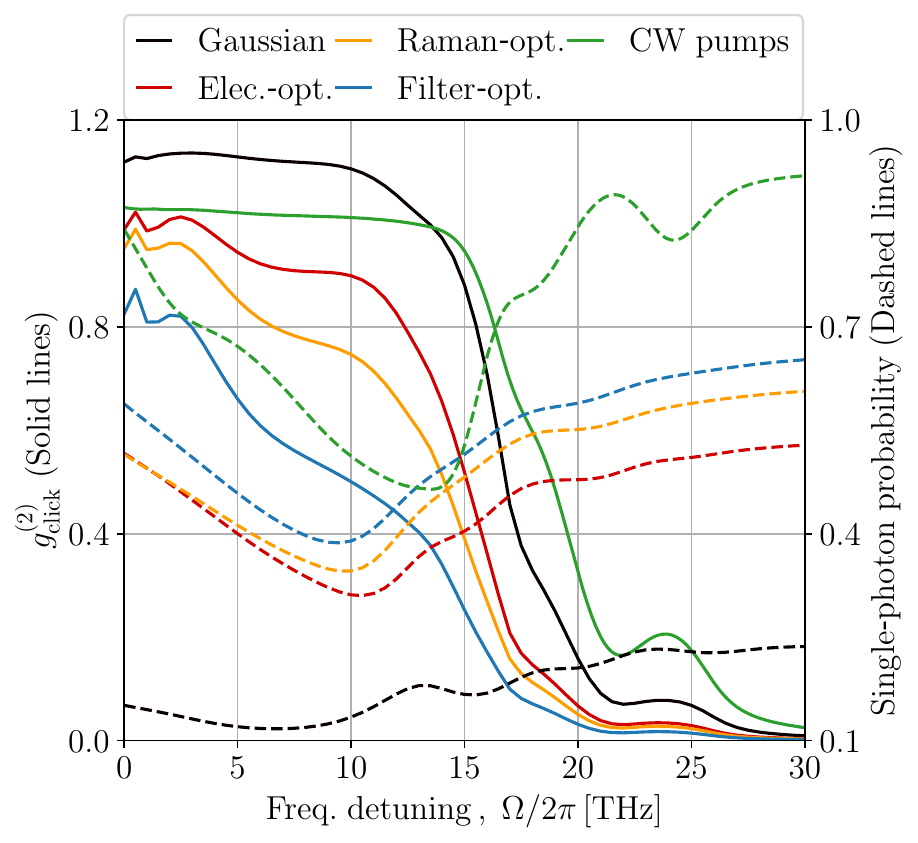}
    \caption{(Left axis) The time-resolved second-order correlation value $g_\mathrm{click}^{(2)}$ for a fiber temperature of 300\,K, and (right axis) the probability of detecting a frequency-converted single-photon, as a function of frequency detuning. These are shown for four choices of input single-photon distributions being a Gaussian centered around $t_\mathrm{in}$, an electronic optimized, a Raman optimized, and a filter-optimized. The green curves illustrates the case with a Gaussian input pulse and CW pumps, from \cref{fig:integrated_g2_CW}(b).}
    \label{fig:pulsed_pump_g2}
\end{figure}

Using the Green functions we can directly determine the second-order correlation value $g_\mathrm{click}^{(2)}$ from \cref{eq:g2_click}.
Figure \ref{fig:pulsed_pump_g2} shows $g_\mathrm{click}^{(2)}$ for a fiber with $\fR = 0.18$, temperature of 300\,K, and a nonlinear interaction strength $\gamma P \ell = 4\pi$ for four choices of input pulses, as a function of the frequency detuning $\Omega$.
For comparison we have included the case with CW pumps from the previous section.
The output quantum states are filtered spectrally, as described by \cref{eq:output_filter}.
The four choices of input pulses represent four methods for optimizing the single-photon shape with respect to the pumps.

The most primitive choice, being the Gaussian, ensures the single-photon has the same shape and duration as the pumps.
This leads to a poor performance due to the complicated phase modulation effects of the pumps.
We take the electronically induced phase-modulation into account, by choosing the input shape to be the first-order Schmidt mode of $G_{\i\s}$ with $\fR = 0$, illustrated by the black line in the inset in \cref{fig:schmidt_mode}.
This improves the $g_\mathrm{click}^{(2)}$-value noticeably.
Including the effects of Raman scattering, by choosing the input Schmidt mode with $\fR =0.18$, as illustrated by the red line in the inset in \cref{fig:schmidt_mode}, improves the value even further.
However, only by taking the filter into account in the Green function, it is possible to achieve the best performance.

For all frequency detunings of interest, the CW case has the worst performance.
This is because the average pump power over the detection window is larger for the CW pumps, which increases the accumulated spontaneous Raman photons.

This preliminary analysis shows that it is necessary to consider all physical effects to achieve the best possible frequency conversion.
However, in a realistic experiment, the single-photon shape is often predetermined, and the pump shapes would therefore have to be optimized with respect to this.
Although this inverse problem is generally difficult to tackle, the present model paves the way for determining the optimal pump shapes by taking the relevant physical effects into account.

\section{Discussion and conclusion}

In this paper, we have presented a general quantum mechanical model which describes the frequency conversion of a quantum state in a medium constrained by Raman scattering.
We have presented the model using a simple configuration with linearly polarized modes, and a single-mode fiber.
However, the generality of the model allows modeling of frequency conversion between complicated spatial and polarization modes by using the appropriate coefficients, which can be found using the expressions provided in the appendix.

Our focus of interest has been
on the impact of Raman scattering on single-photon applications of Bragg scattering.
Therefore, the appropriate figure of merit is the second-order correlation function, for which we determined an expression in terms of Green functions.

Using CW pumps we found analytical expressions of the Green functions in the frequency domain, from which we have shown that it is possible to mitigate the impact of Raman scattering by cooling the fiber, adjusting the frequency detuning, or choosing a crosspolarized field configuration.
The results shows that it is not possible to achieve a reasonable $g_\mathrm{click}^{(2)}$-value within the Raman bandwidth at room temperature, and a frequency beyond the Raman bandwidth is experimentally challenging due to increasing phase-matching sensitivity~\cite{guasoniIntermodalFourWaveMixing2017}.
In order to achieve a $g_\mathrm{click}^{(2)}$-value below 0.05 at the dominant Raman peak, $\Omega = 13\,\mathrm{THz}$, with a copolarized configuration, a temperature of $\sim135\,\mathrm{K}$ is necessary, while a temperature of $\sim253\,\mathrm{K}$ is sufficient with a crosspolarized configuration.

Finally, we have developed a numerical split-step scheme for determining the Green functions in a pulsed pump configuration.
This opens the parameter space considerably, and a careful analysis of a specific system is necessary for designing the optimal setup.
However, with the preliminary analysis, we have shown that it is possible to optimize the $g_\mathrm{click}^{(2)}$-value considerably, by taking the relevant physical aspects of an experimental setup into account.

{\bf Acknowledgements}
This work was financially supported by the Danish Council for Independent Research (DFF) (ref. 1032-00460B) and by the Innovation Fund Denmark (IFD) (ref. 2079-0040B). In addition, we would like to acknowledge Rodrigo Silva, funded by the European Union’s Horizon Europe Research and Innovation Programme under the Marie Sklodowska-Curie Grant Agreement No. 101072409, for fruitful discussions.

\section{Appendices}
\subsection{Nonlinear coefficients}\label{app:nonlinear_coef}
Here we give a justification for using the vectorial model described by Lin et al.~\cite{linPhotonpairGenerationOptical2007}, to model quantum fields quantized in arbitrary spatial or polarization modes.
We start by considering the full-vectorial Hamiltonian describing the electronic interaction, and Raman scattering~\cite{drummondQuantumTheoryFiberOptics1996a}
\begin{subequations}
\begin{align}
    \hat{H}_\e &= \int \ud^3\mathbf{x} \frac{1}{4\varepsilon^3}\sum_{ijkl} \chi_{ijkl}^{(3)} :\hat{D}_i\hat{D}_j \hat{D}_k \hat{D}_l:\\
    \hat{H}_\mathrm{R} &= \int \ud^3\mathbf{x} \int_0^\infty\ud \omega \sum_{ij} \mathcal{R}_{ij}(\omega) :\hat{D}_i\hat{D}_j\hat{q}(\omega,\mathbf{x}):
\end{align}
\end{subequations}
where $\chi_{ijkl}^{(3)}$ is the third-order susceptibility tensor, $\varepsilon$ is the material permitivity, $\mathcal{R}_{ij}$ is the Raman response tensor, $\hat{q}$ is the bosonic phonon operator, $\hat{D}_i$ is the $i$'th vector component of the displacement field, and we have defined $\mathbf{x} = (x,y,z)$.
In contrast to previous work dealing with Raman scattering, we choose the phonon operator to be localized in all three spatial coordinates $\mathbf{x}$.
This gives rise to a multimode description of Raman scattering, which is necessary to construct a model for intermodal FWM-BS.
We perform a mode-field expansion of the fields, to find the Hamiltonian in terms of the creation and annhilation operators~\cite{quesadaPhotonPairsNonlinear2022}
\begin{subequations}
\begin{align}
    \hat{\mathbf{D}}(\mathbf{x},t) &= \sum_m \sqrt{\frac{\hbar\varepsilon\omega_m}{2v_m}}\left(\hat{A}_m(z,t) \mathbf{d}^{(m)}(\mathbf{r})\e^{i\beta_{m0}z - i\omega_m t} + \hc\right)\\
    \hat{q}(\omega,\mathbf{x}) &\propto \hat{b}(\omega,\mathbf{x}) + \hat{b}^\dagger(\omega,\mathbf{x})
\end{align}
\end{subequations}
where $\omega_m$ and $v_m$ is the center frequency and group velocity of field $m$, respectively.
The mode profiles are normalized $\int\ud^2\mathbf{r}|\mathbf{d}^{(m)}(\mathbf{r})|^2 = 1$, and are not necessarily orthogonal.
Inserting this into the electronic Hamiltonian, and keeping only photon-number-conserving terms, we find the mode-field Hamiltonian for the electronic interactions
\begin{equation}
    \hat{H}_\e = \hbar \int\ud z \left(2 - \tfrac{3}{2}\delta_{abcd}\right)\Gamma_{abcd} \hat{A}_a^\dagger\hat{A}_b^\dagger\hat{A}_c\hat{A}_d
\end{equation}
where we have defined the nonlinear coefficient
\begin{align}
    \Gamma_{abcd} &= \frac{3\hbar}{4\varepsilon}\sqrt{\frac{\omega_a\omega_b\omega_c\omega_d}{v_av_bv_cv_d}}\sum_{ijkl}\chi_{ijkl}^{(3)}O_{ijkl}^{(abcd)}\\
    O_{ijkl}^{(abcd)} &= \int \ud^2\mathbf{r} d_i^{(a)*}(\mathbf{r})d_j^{(b)*}(\mathbf{r})d_k^{(c)}(\mathbf{r})d_l^{(d)}(\mathbf{r})
\end{align}
where $O_{ijkl}^{(abcd)}$ is the inverse mode area.
Similarly, we find the Raman scattering Hamiltonian in terms of the creation and anhilation operators
\begin{align}
    \hat{H}_\mathrm{R} = \hbar &\int \ud^3 \mathbf{x} \int_0^\infty\ud\omega\notag\\
    \times&\sum_{ij}\mathcal{R}_{ij}(\omega) d_i^{(a)*}(\mathbf{r})d_j^{(b)}(\mathbf{r}) \hat{q}(\mathbf{x},\omega)\hat{A}_a^\dagger\hat{A}_b\,.
\end{align}
Following the procedure outlined by Drummond~\cite{drummondQuantumTheoryFiberOptics1996a} and Lin et al.~\cite{linPhotonpairGenerationOptical2007} we arrive at the evolution equation of the quantum fields
\begin{align}
    \partial_z \A_m &= - \beta_{1m} \partial_t \A_m -\frac{i}{2}\beta_{2m} \partial_t^2 \A_m\notag\\
    &+ i\sum_{nkl}\Gamma_{mnkl} (2-\delta_{mn})\Ad_n\A_k\A_l+ i\sum_n \m_{mn}\A_n \notag \\
    &+ i\sum_{nkl}\A_n \int \ud t' f_{mnkl}(t-t') \Ad_k(t')\Ad_l(t')
\end{align}
where the noise operator, describing spontaneous Raman scattering is given by
\begin{align}
    \m_{mn}(z,t) &= \int_0^\infty \ud\omega \mathcal{R}_{ij}(\omega) \int \ud^2\vec{r} d_i^{(n)*}(\vec{r})d_j^{(m)}(\vec{r}) \notag \\
    &\quad\times \left [\b(\vec{x},\omega)\e^{-i\omega t} + \bd(\vec{x},\omega)\e^{i\omega t}\right]\e^{i(\omega_m - \omega_n) t}
\end{align}
and the Raman response function is found to be
\begin{subequations}
\begin{align}
    f_{abcd}(t) = 2 \Theta(t)& \e^{i(\omega_a - \omega_b) t}\int_0^\infty\ud \omega \sin(\omega t) \mathrm{Im}\{f_{abcd}(\omega)\}\\
    \mathrm{Im}\{f_{abcd}(\omega)\} &= \sum_{ijkl} \mathcal{R}_{ij}(\omega)\mathcal{R}_{kl}(\omega)O_{ijkl}^{(abcd)}
\end{align}
\end{subequations}
The response function and the phonon noise oscillates at the frequency difference $\omega_a - \omega_b$.
When the field are chosen to be the four fields described in \cref{fig:pol_config} this corresponds to the frequency detuning $\Omega$.
When introducing the semi-classical approximation the nonlinear coefficient becomes,
\begin{equation}
    \gamma_{abcd} = \frac{\Gamma_{abcd}}{\hbar\sqrt{\omega_a\omega_c}}
\end{equation}
where $a$ and $c$ are the pump fields.
In the Raman response functions the coefficient $(\hbar\sqrt{\omega_a\omega_c})^{-1}$ is absorbed into the Raman response tensor $\mathcal{R}_{ij}$.

\subsection{Second-order correlation function}\label{app:g2}
We want to find an expression for the second-order correlation function in terms of the Green functions.
To simplify the evaluation, we express the output idler operator in terms of two operators, representing the optical, and phononic part respectively
\begin{equation}
    \a_\i(\ell,t) = \AA(t) + i \MM(t).
\end{equation}
Using that $\mean{\MM} = 0$, we determine the four point correlation for a single-photon input state, which thereby yields
\begin{align}
    &\mean{\ad_\i(t_1)\ad_\i(t_2)\a_\i(t_2) \a_\i(t_1)} = |\psi_\mathrm{out}(t_1)|^2 E(t_2,t_2)\\
    &\quad + |\psi_\mathrm{out}(t_2)|^2 E(t_1,t_1) + \psi_\mathrm{out}^*(t_1)\psi_\mathrm{out}(t_2) E(t_2,t_1)\notag\\
    &\quad + \psi_\mathrm{out}^*(t_2)\psi_\mathrm{out}(t_1) E(t_1,t_2)\notag\\
    &\quad + \mean{\MM^\dagger(t_1)\MM^\dagger(t_2)\MM(t_2)\MM(t_1)}\notag
\end{align}
where we have used that the optical four-point correlation function, and the correlation functions odd number of creation or annihilation operators evaluate to zero, and we have defined the expectation values
\begin{subequations}
\begin{align}
    E(t_1,t_2) &= \mean{\MM^\dagger(t_1)\MM(t_2)},\\
    \psi_\mathrm{out}^*(t_1)\psi_\mathrm{out}(t_2) &= \mean{\AA^\dagger(t_1)\AA(t_2)}.
\end{align}
\end{subequations}
The phononic four-point correlation function can be factorized since the phonons bath exhibits thermal statistics:
\begin{align}
    \mean{\MM^\dagger(t_1)\MM^\dagger(t_2)\MM(t_2)\MM(t_1)} &= E(t_1,t_1)E(t_2,t_2)\\
    &+ E(t_1,t_2)E(t_2,t_1)\notag
\end{align}
and thereby we obtain the numerator of \cref{eq:general_g2}.







\bibliography{thesis}

\end{document}